\documentclass[preprint,showpacs,amsmath,amssymb]{revtex4}

\usepackage{graphicx}
\usepackage{dcolumn}
\usepackage{bm}

\begin{document}
\newcommand{\kp}{{\bf k$\cdot$p}\ }

\title{Nernst-Ettingshausen effect at the trivial-nontrivial band ordering in topological crystalline insulator Pb$_{1-x}$Sn$_x$Se}

\author{K. Dybko}
\email{Krzysztof.Dybko@ifpan.edu.pl}
\author{P. Pfeffer}
\author{M. Szot}
\author{A. Szczerbakow}
 \author{A. Reszka}
 \author{T. Story}
 \author {W. Zawadzki}
 \affiliation{Institute of Physics, Polish Academy of Sciences\ \\
 Al.Lotnikow 32/46, 02--668 Warsaw, Poland}
\date{\today}
\begin{abstract}

The transverse Nernst Ettingshausen ($N-E$) effect and electron mobility in Pb$_{1-x}$Sn$_x$Se alloys are studied experimentally and theoretically as functions of temperature and chemical composition in the vicinity of vanishing energy gap $E_g$. The study is motivated by the recent discovery that, by lowering the temperature, one can change the band ordering from trivial to nontrivial one in which the topological crystalline insulator states appear at the surface. Our work presents several new aspects. It is shown experimentally and theoretically that the bulk $N-E$ effect has a maximum when the energy gap $E_g$ of the mixed crystal goes through zero value. This result contradicts the claim made in the literature that the $N-E$ effect changes sign when the gap vanishes. We successfully describe $dc$ transport effects in the situation of extreme band's nonparabolicity which, to the best of our knowledge,  has never been tried before. A situation is reached in which both 
 two-dimensional bands (topological surface states)    and three-dimensional bands are linear in  electron \textbf{k} vector. Various scattering modes and their contribution to transport phenomena in Pb$_{1-x}$Sn$_x$Se are analyzed. As the energy gap goes through zero, some transport integrals have a singular (nonphysical) behaviour and we demonstrate how to deal with this  problem by introducing damping.
\end{abstract}

\maketitle
\section{INTRODUCTION}
Narrow gap semiconductors have been for many years subject of intense experimental and theoretical studies in view of their interesting properties and important applications \cite{nimtz,khokh,spring}.    In recent years they have become once again a source of excitement due to the discovery of topological insulators  \cite{Molenkamp}. The topological boundary   states have been observed in bulk compounds Bi$_2$Se$_3$, Bi$_2$Te$_3$, bulk alloys Bi$_{1-x}$Sb$_x$\cite{HasanMoore}, two-dimensional quantum wells of HgTe/Hg$_{1-x}$Cd$_x$Te \cite{Molenkamp} and, most recently, in bulk Pb$_{1-x}$Sn$_x$Se alloys \cite{dzia}, bulk SnTe \cite{Tanaka} and bulk Pb$_{1-x}$Sn$_x$Te alloys \cite{XU}. The latter were called topological crystalline insulators (TCI), because, in contrast to canonical topological insulators,  specific crystalline symmetries warrant the topological protection of their metallic surface states \cite{fu,hsi,andoFU}. Since the IV-VI lead chalcogenides are characterized by strong temperature dependence of their band structures, it is possible to reach proper band ordering for a suitably chosen chemical composition $x$ by varying the temperature. For the Pb$_{1-x}$Sn$_x$Se system at high temperatures the band ordering is called trivial ($L_6^-$ band above $L_6^+$ band) and no TCI state occurs. As the temperature is lowered for the properly chosen $x$, one can reach the vanishing band gap and then arrive at the inverted nontrivial ordering ($L_6^+$ band above $L_6^-$ band) in which TCI state can occur. Such transition was demonstrated with the use of angle-resolved photoemission spectroscopy in Ref. \cite{dzia}. Generally speaking, the character of topological insulators is manifested in both their surface and bulk properties. 

It has been recognized from the sixties that in the alloys Pb$_{1-x}$Sn$_x$Se one can reach vanishing energy gap by changing the temperature. An early demonstration of this possibility was provided by Strauss \cite{stra} who used for this purpose the optical transmission measuring the gap on both sides of the band ordering. Also laser emission proved useful in this respect \cite{har}. It was shown that the electric resistivity and the Hall coefficient depend on the band ordering and can be used to obtain information on the transition temperature even if the Fermi energy is quite high in either the valence or conduction band \cite{dix}. Quite recently, optical response with appropriate analysis was used to determine the band inversion temperature for Pb$_{0.77}$Sn$_{0.23}$Se \cite{reij,xi,Anand}. 

In recent years the thermomagnetic Nernst-Ettingshausen effect experiences a real revival in investigations of semiconductors \cite{liang,evola}, graphene \cite{peng,chec} and high-T$_c$ superconductors \cite{yayu,bech}.
In the present work we undertake a bulk transport study of  PbSe and Pb$_{1-x}$Sn$_x$Se system for $0.25 \leq x < 0.39$. The above ternary alloys  can reach the vanishing band gap in the available temperature range. We investigate experimentally and theoretically the conduction electron density, mobility and the $N-E$ effect. Since our study concentrates on small gap values, particular features of the band structure are crucial and it turns out that the description of transport effects for the vanishing gap poses nontrivial theoretical problems. In fact, although we describe bulk properties, as the gap goes to zero we
deal with linear energy bands which is a completely new situation for the
transport theory.
   We demonstrate for four Pb$_{1-x}$Sn$_x$Se samples that, as the gap goes to zero, the $N-E$ effect reaches a maximum. This contradicts the claims made in the literature, see Ref. \cite{liang}. All in all, our results confirm the conclusions of Ref. \cite{dzia} concerning the transition from trivial to nontrivial band ordering as a function of temperature. Our analysis of scattering modes and their relative importance in Pb$_{1-x}$Sn$_x$Se near the trivial-nontrivial transition of the band ordering will help further investigations of topological crystalline insulators and other topological materials.

\section{EXPERIMENT}
Single crystals of Pb$_{1-x}$Sn$_x$Se (0$\leq x \leq$ 0.39) were grown by self-selecting vapour growth technique \cite{szcz1,szcz2}. Owing to the peculiarity of the method, where near equilibrium thermodynamic conditions are kept, we have obtained high quality compositionally uniform large monocrystals with natural (001) facets (typical dimensions: 1$\times$1$\times$1 cm$^3$).
Crystal compositions were determined by Energy Dispersive X-ray Spectroscopy offered by Scanning Electron Microscope Hitachi SU-70. Molar fractions were taken as averages from the scan covering 1$\times$ 1.5 mm$^2$ of the sample surface. The accuracy of determination of the chemical composition is better than 0.005 molar fraction.

Samples for measurements were cleaved with razor blade along (001) planes in the form of rectangular parallelepipeds, with dimensions 1.5$\times$3$\times$10 mm$^3$. 
All samples exhibited metallic behaviour of resistivity with almost temperature independent carrier concentration. 
The Hall effect measured electron density in PbSe  to be 7$\cdot10^{18}$ cm$^{-3}$ and the mobility 35000 cm$^2$/Vs at liquid helium temperature. The corresponding values  for ternary compounds Pb$_{1-x}$Sn$_x$Se  are included in Table 1.  For thermoelectric measurements, the samples were thermally anchored with silver epoxy to copper cold finger of continuous flow helium cryostat. A small SMD resistor was glued to the free end of the sample for use as a heater.
Copper potential leads were attached across the sample with silver paint. Temperature and temperature gradient were measured with two calibrated subminiature GaAlAs diodes by Lake Shore glued to the sample. For all temperatures, the temperature gradient was kept within 2-5 percent of the average temperature.
An external magnetic field at given temperature, with proper temperature gradient steady developed, was swept from -0.5 to 0.5 Tesla using standard resistive magnet. The magnetic field was directed perpendicularly to the sample and the temperature gradient. In this configuration the transverse $N-E$ electric field develops across the sample:
\begin{equation}
\bm{E}_{\rm N-E}= P_{\rm N-E } \cdot \bm{{\nabla}} T \times \bm{B}\;\;,
\end{equation}
where P$_{\rm N-E}$ is the $N-E$ coefficient, $T$ is the temperature and $\bf{B}$ is the magnetic field.
The transverse $N-E$ effect is a thermoelectric analogue of the Hall effect. The potential difference V$_{\rm N-E}$=w$\cdot$E$_{\rm N-E}$ (w is sample's width) was measured by Keithley nanovoltmeter 2182a. The potential was linearly dependent on the magnetic field for all applied fields. Main uncertainties in the absolute P$_{\rm N-E} $ data are systematic and arise mainly from finite size of the thermometers contacts to the sample. We estimate their contribution to the uncertainty of the temperature gradient to be less than 15 percent. All other possible errors are small in comparison.

\section{THEORY}
The conduction band of PbSe and Pb$_{1-x}$Sn$_x$Se alloys consists of four ellipsoids of revolution with minima at the four L points of the Brillouin zone. As a consequence, one deals with the longitudinal $m^*_l$ and transverse $m^*_t$ effective masses, the anisotropy in PbSe at 0 K is $m^*_l/m^*_t$= 1.7. In the description of electron scattering one deals with the density-of-states mass $m^*_d = (m^*_l {m^*_t}^2)^{1/3}$ and in the description of mobility with the conductivity mass $m^*_c = 3/({m^*_l}^{-1}+2{m^*_t}^{-1})$. In PbSe and Pb$_{1-x}$Sn$_x$Se these masses differ little from each other, so we use an approximate spherical value of $m^*_0 =0.048 m_0$ at 0 K at the band edge ($m_0$ is free electron's mass). This means that we use the standard dispersion relation for the two-band Kane model (the zero of energy is chosen at the band edge)
\begin{equation}
{\cal E}= \frac{-E_g}{2}+\left[\left(\frac{E_g}{2}\right)^2 + E_g\frac{\hbar^2k^2}{2m^*_0}\right]^{1/2}\;\;,
\end{equation}
This gives the energy-dependent mass, see \cite{wz}
\begin{equation}
m^*({\cal E}) = m^*_0\left(1+2\frac{\cal E}{E_g}\right)\;\;,
\end{equation}
It is seen from Eqs. (2) and (3) that, as the gap vanishes, we deal with
the linear energy dispersion E(k) i.e. with, so called,  massless Dirac fermions.
In principle the {\kp} theory is valid at $T$ = 0, but here it is known to work also for higher temperatures. Using experimental data on the composition dependence of the band gap of Pb$_{1-x}$Sn$_x$Se at various temperatures, see Refs. \cite{stra,nimtz,woj}, we establish the following phenomenological dependence of $E_g$ $\rm{(meV)}$ on $x$ and $T$
\begin{equation}
E_g(T, x) = 126-802 \cdot x+[400+0.258 \cdot T^2]^{1/2} \;\;,
\end{equation}
In the two-band {\kp} model the effective mass is proportional to the gap and inversely proportional to the matrix element of momentum squared. The temperature dependence of the mass at the band edge of Pb$_{1-x}$Sn$_x$Se is
\begin{equation}
\frac{m^*_0}{m_0} = \frac{E_g \rm{(eV)}}{3.0546+1.8626 \cdot 10^{-3}\cdot T}\;\;.
\end{equation}
In order to consistently   describe transport effects we use the formalism developed for nonparabolic energy bands, see \cite{wz}. According to this scheme, statistical and transport quantities are given by the integrals
\begin{equation}
<A> = \int_0^{\infty}\left(-\frac{\partial f_0({\cal E})}{\partial \cal E}\right)A({\cal E})k^3 d{\cal E}\;\;,
\end{equation}
where $f_0$ is the Fermi-Dirac distribution function depending on the Fermi level $E_f$. Since the derivative $\partial f_0({\cal E})/ \partial \cal E$ does not vanish only in the limited range of energies around the Fermi level, the integrals (6) are not difficult to compute. If the electron gas is strongly degenerate, there is $\partial f_0({\cal E})/ \partial {\cal E} = - \delta({\cal E}-E_f)$ and the integrals (6) are equal to the integrands taken at the Fermi energy. The free electron density in the band is
\begin{equation}
N = \frac{N_v}{3\pi^2}\int_0^{\infty}\left(-\frac{\partial f_0({\cal E})}{\partial \cal E}\right)k^3d{\cal E} = \frac{N_v}{3\pi^2}<1>\;\;,
\end{equation}
where $N_v$ = 4 is the number of equivalent ellipsoids. The electron mobility is
$\mu({\cal E}) = e \tau({\cal E})/m^*({\cal E})$\;,
in which $\tau$ is the relaxation time and $m^*({\cal E})$ is given by Eq. (3). The electric conductivity is $\sigma  = e N \overline{\mu}$, where the average electron mobility is
\begin{equation}
\overline{\mu} = \frac{<\mu>}{<1>}\;\;.
\end{equation}
An experimental determination of free electron density is nontrivial for our samples since it involves the Hall scattering factor
\begin{equation}
A_r = \frac{<\mu^2><1>}{<{\mu}>^2} = \frac{\overline{\mu^2}}{(\overline{\mu})^2}
\end{equation}
which for materials with vanishing energy gap should be carefully evaluated, see below. The $N-E$ effect, which is the main subject of our interest, is described in Ref. \cite{wz}
\begin{equation}
P_{\rm N-E} = -\frac{k_B}{|e| c}A_r {\overline \mu}\left(\frac{<z\mu^2>}{<\mu^2>} - \frac{<z\mu>}{<\mu>}\right)\;\;,
\end{equation}
where: $k_B$ - Boltzmann constant and $z={\cal E}/k_BT$. For the complete degeneracy of electron gas the difference in the parenthesis is zero and the $N-E$ effect vanishes.

In order to describe electron scattering mechanisms one needs to know the electron wave functions for the conduction band. These are taken in the form of true Bloch states with the periodic amplitudes depending on the pseudo-momentum $\hbar \textbf{k}$ and energy, see \cite{rav}. Because of the narrow energy gap and fairly strong spin-orbit interaction the wave functions are in general linear combinations of conduction and valence periodic states as well as mixtures of spin-up and spin-down components. The wave functions for the conduction band in the spherical approximation are, cf. \cite{rav}
 \ \\
 \begin{equation}
\Psi^+_c = \left[\sqrt{1-L} b Z + \sqrt{L}\frac{k_z}{k} iR\right]\uparrow - \left[\sqrt{1-L} d X_+ + \sqrt{L}\frac{k_-}{k} iR\right]\downarrow\;\;,
\end{equation}
\ \\
\begin{equation}
\Psi^-_c = \left[\sqrt{1-L} b Z + \sqrt{L}\frac{k_z}{k} iR\right]\downarrow + \left[\sqrt{1-L} d X_- + \sqrt{L}\frac{k_+}{k} iR\right]\uparrow\;\;,
\end{equation}
\ \\
where $L={\cal E}/(2{\cal E}+E_g)$ and $k_{\pm} = k_x \pm ik_y$, while Z, X$_{\pm}$ = (X$\pm$ iY)/$\sqrt{2}$ and R denote the periodic amplitudes of Luttinger-Kohn functions taken at the L points of the Brillouin zone. The normalization coefficients satisfy the condition $b^2+d^2$=1. Plus and minus signs in the overscripts of the wave functions symbolize the effective spins up and down, respectively, while the arrows on the RHS symbolize spin-up and spin-down functions quantized on the \textbf{z} direction. The electron energies $\cal E$ are counted from  the conduction band edge.

The scattering probabilities have been calculated for various scattering modes by computing matrix elements of corresponding perturbing potentials. The total scattering probability is a sum of separate probabilities, which amounts to calculating the total relaxation time according the well known formula
\begin{equation}
{\tau}^{-1} = \sum_i {\tau}_i^{-1}\;\;,
\end{equation}
where $\tau_i$ describe relaxation times for specific (independent) modes.
Below we enumerate the relevant scattering modes, quote the involved material parameters and mention importance of various modes for the electron mobility and $N-E$ effect in Pb$_{1-x}$Sn$_x$Se for the chemical compositions $0.25 \leq x \leq 0.39$ of our interest. The description includes band's nonparabolicity in the energies and wave functions, as well as screening of long-range interactions by the electron gas.

1) Polar scattering by optic phonons (OP). This mode is important at higher temperatures. It is determined by the static and high-frequency dielectric constants ${\varepsilon}_0$ =234 and ${\varepsilon}_{\infty}$ = 28.7 at $T$ = 0. Their temperature variations can be found in \cite{nimtz, pre}. The mode is characterized by nonelastic scattering processes which complicates its description in relaxation time approximation.  This mode contains no adjustable parameters.
Taking into account the screening of the interaction by the electron gas one obtains for the relaxation time, cf. \cite{rav}
\ \\
\begin{equation}
\frac{1}{{\tau}_{pol}} = \frac{2 k_BT e^2}{\hbar}(\frac{1}{{\varepsilon}_{\infty}}-\frac{1}{{\varepsilon}_0})\frac{d k}{d \cal E}F_{pol}\;,
\end{equation}
\ \\
where
\ \\
\begin{equation}
F_{pol} = \left[1-\frac{ln(1+\rho_{\infty})}{\rho_{\infty}}\right]
-2L(1-L)\left[1-\frac{2}{\rho_{\infty}}+2\frac{ln(1+\rho_{\infty})}{{\rho_{\infty}}^2}\right]\;\;,
\end{equation}
\ \\
in which $\rho_{\infty} = 4k^2{\lambda_{\infty}}^2$, where $\lambda_{\infty}$ is the screening length for ${\varepsilon}_{\infty}$.

2) Nonpolar scattering by optic phonons (NOP). This mode is determined by electron-optic-phonon deformation potential interaction. Due to the interband {\kp} mixing it involves both the conduction and valence deformation potential constants, $E_{np}^c$ and $E_{np}^v$, which are treated as adjustable parameters, cf. \cite{zaya}. We assumed that $E_{np}^c$ and $E_{np}^v$ are equal. A correction due to the nonelasticity is  included, see \cite{morg}.
The relaxation time for this mode is
\ \\
\begin{equation}
\frac{1}{{\tau}_{np}} = (E^c_{np})^2\frac{k_BT\hbar}{\pi (\hbar \omega_{op})^2 a^2_0}\frac{\partial k}{\partial{\cal E}} k^2T_{cor} F_{np} \;,
\end{equation}
\ \\
where
\ \\
\begin{equation}
F_{np} = \left[1-L\left(1-\frac{E_{np}^v}{E_{np}^c}\right)\right]^2-L(1-L)\frac{8}{3}\frac{E_{np}^v}{E_{np}^c}\;\;.
\end{equation}
\ \\
Here $\hbar\omega_{op}$ is the energy of the optical phonon and $a_0$ is the lattice constant. The correcting term $T_{cor}=(\exp z-1)^2/(z^2 \exp z)$, where $z={\cal E}/{k_BT}$.

3) Scattering by acoustic phonons (AC). It involves conduction and valence acoustic deformation potentials $E_{ac}^c$ and $E_{ac}^v$ treated as adjustable parameters. We assumed that $E_{ac}^c$ and $E_{ac}^v$ are equal - a good approximation for mirror-like conduction band - valence band symmetry in IV-VI semiconductors. The relaxation time for this mode is
\ \\
\begin{equation}
\frac{1}{{\tau}_{ac}} = (E^c_{ac})^2\frac{k_BT}{\pi \hbar v^2_{av} \varrho} \frac{\partial k}{\partial{\cal E}} k^2F_{ac}\;,
\end{equation}
\ \\
\begin{equation}
F_{ac} = \left[1-L\left(1-\frac{E_{ac}^v}{E_{ac}^c}\right)\right]^2-L(1-L)\frac{8}{3}\frac{E_{ac}^v}{E_{ac}^c}\;\;,
\end{equation}
\ \\
where
$v_{av}$ is the averaged sound velocity and $\varrho$ is the crystal density.
\ \\

4) Scattering by ionized defects is due to electrostatic interaction between electrons and charged defects in the crystal. In  Pb$_{1-x}$Sn$_x$Se each native defect furnishes two free electrons (Se vacancies) or two free holes (metal vacancies) \cite{nimtz}.
In general, the defect potential is of the form V=V$_C$ + V$_{sr}$, where V$_C$ is the Coulomb interaction and V$_{sr}$ symbolizes the short range interaction related to size of the defect. The Coulomb interaction is negligible in lead chalcogenides due to very high value of the static dielectric constant ${\varepsilon}_0$. Thus, one is left with the short-range contribution to electron scattering. The relaxation time for this mode is, see \cite{lit}
\ \\
\begin{equation}
\frac{1}{{\tau}_{sr}} = \frac{A^2 N_{d}F_{sr}}{2\pi \hbar}\frac{d k}{d \cal E}k^2\;,
\end{equation}
\ \\
\begin{equation}
F_{sr} = \left[1-L\left(1-\frac{B}{A}\right)\right]^2-L(1-L)\frac{8}{3}\frac{B}{A}\;\;,
\end{equation}
\ \\
where $N_d$ is the concentration of ionized defects, while A = $<R|V_{sr}|R>$ and B=$<X|V_{sr}|X>$ are the matrix elements of the short-range potential V$_{sr}$ for the conduction and valence bands, respectively.
The elements A and B are treated as adjustable parameters and were assumed to be equal.

5) Alloy disorder scattering (AD), that appears only in the ternary alloys, is due to the fact that the $V_{Pb}$ and $V_{Sn}$ atomic potentials are not the same. This results in perturbations of crystal periodicity and, consequently, in electron scattering, see \cite{koss,taki}. The disorder scattering is important at high values of $x$ and low temperatures. The relaxation time for this mode is, see \cite{taki}
\ \\
\begin{equation}
\frac{1}{{\tau}_{ad}} = (U^c_{ad})^2\frac{4x(1-x)}{\pi \hbar \Omega}\frac{\partial k}{\partial{\cal E}} k^2F_{ad} \;,
\end{equation}
where $U^c_{ad}$ is the matrix element of the potential difference $V_{Pb} - V_{Sn}$ for the conduction band and $\Omega$ is the volume of unit cell. Further
\ \\
\begin{equation}
F_{ad} = \left[1-L\left(1-\frac{U_{ad}^v}{U_{ad}^c}\right)\right]^2-L(1-L)\frac{8}{3}\frac{U_{ad}^v}{U_{ad}^c}\;\;,
\end{equation}
\ \\
where $U_{ad}^v$ is the matrix element of the potential difference $V_{Pb-Sn}$ for the valence band. According to the theory, see \cite{koss}, there exists a relation
\ \\
\begin{equation}
E_g(SnSe)=E_g(PbSe)+\frac{(U^c_{ad}-U^v_{ad})}{\Omega}\;.
\end{equation}
\ \\
Knowing the gaps of both materials and fitting the value of $U_{ad}^c$ one automatically obtains the value of $U_{ad}^v$. 

6) Scattering by charged dislocations (DIS). This mode is due to repulsive interaction of electrons with dislocation lines which, forming acceptor centers, attract conduction electrons and become negatively charged, see \cite{podo}. The mode depends strongly on dislocation density (which is an adjustable parameter) and rather weakly on the temperature. The relaxation time for this mode is
\ \\
\begin{equation}
\frac{1}{{\tau}_{dis}}=\frac{f^2e^4\lambda^4_0 N_{dis}{m^*_0}(1+2{\cal E}/E_g)}{\hbar^3\varepsilon_0^2(1+k^2_\perp\lambda_0^2)^{3/2}a^2}\;\;,
\end{equation}
where $f$ is the fraction of filled traps, $a$ is the lattice constant, $\lambda_0$ is the screening length for ${\varepsilon}_0$ and $N_{dis}$ is the dislocation density.

In Table 1 we quote the employed values of material parameters. After the adjustment they are kept the same for all four samples. The density of dislocations $N_{dis}$ and the damping constant $G$ are adjusted for each sample separately.

Trying to compare experimental transport data with the theory one has to connect the measured quantities with the calculated ones. To work out this question we proceed in the following way. We begin with the experimentally determined Hall constant $R_H$ for a given temperature $T$ from which we determine the Hall electron density $N_H = 1/e R_H$. The real free-electron density is $N = A_r N_H$, where $A_r$ is the Hall scattering factor. From the given value of $N_H$ we compute the first value of the Fermi energy $E_f$ and use $E_f$ to compute the first value of mobility $\overline{\mu}$ using Eq. (8) with the above mentioned  scattering modes, and finally the first value of $A_r$ given by Eq. (9). Employing this value of $A_r$ we determine the first value of $N$ which concludes the first round of iteration. For this value of $N$ we compute the second value of $E_f$, etc, all the way to the second value of $N$. We terminate the iteration when the (n+1)th value of $N$ is practically the same as the nth. This procedure determines the final values of $N$, $E_f$, $A_r$ and $\overline{\mu}$ for the given temperature. These values can then be used to calculate the $N-E$ coefficient given by Eq. (10). The adjustable parameters mentioned above are chosen to obtain an overall best agreement between the experiment and theory for $\overline{\mu}$ and $P_{\rm N-E}$ for our four samples at all available temperatures.

When employing the above scheme we have found that the computed quantity containing the integral $<{\mu}^2>$, that is $A_r$ and $P_{\rm N-E}$, (see Eqs. (9) and (10)), have a strong and narrow peak at the temperature $T_c$ for which $E_g = 0$. Such peak has no physical meaning and, as it can be seen in Fig. 1 for $P_{\rm N-E}$, it is not observed experimentally. Also, there is no reason to expect a sharp peak of the scattering factor $A_r$ because it would lead to a sharp peak of the free electron density $N$ without a  physical reason.  However, since our Pb$_{1-x}$Sn$_x$Se samples are certainly not homogeneous, i.e. they have somewhat different chemical compositions $x$ at various parts, one can not expect to have $E_g = 0$ in the whole sample at one temperature $T$. We simulate this nonhomogeneity by using the well known mathematical measure to avoid singularities in resonances. Thus, we introduce damping by replacing in all formulas the value of $E_g$ by $E_g+iG$ and take real values of the resulting expressions.
By adjusting the damping constant $G$ we can bring both the scattering factor $A_r$ and the $N-E$ coefficient into a reasonable and experimentally observed behaviour, see below.

\section{RESULTS AND DISCUSSION}
Figure 1 shows our main experimental results on the $N-E$ effect in PbSe and four Pb$_{1-x}$Sn$_x$Se samples in which the gaps $E_g$ in  Pb$_{1-x}$Sn$_x$Se alloys can go through zero as functions of temperature. Our absolute experimental values of $P_{\rm N-E}$ are in general agreement with those measured on other narrow-gap semiconducting materials \cite{jetp,evola}. The vertical thin arrows indicate critical temperatures $T_c$ which, according to the dependence (4), correspond to $E_g$ = 0 for the indicated chemical compositions. It can be seen that the maximum of $P_{\rm N-E}$ for each sample corresponds to the critical temperature $T_c$. In other words, a maximum of $P_{\rm N-E}$ indicates that the gap goes through zero value. On the other hand, in PbSe there is no possibility of reaching vanishing gap and one observes no maximum. The observed values of $T_c$, are compared to the line drawn on the basis of $E_g(T, x)$ given in Eq. (4) and presented in Fig. 2. The agreement between the two is very good.

This result contradicts the claim made in Ref. \cite{liang} that the temperature at which the band gap vanishes corresponds to the change of sign of $P_{\rm N-E}$. In particular, in the samples with $x$ = 0.25 and $x$ = 0.277 the coefficient $P_{\rm N-E}$ does not change sign at all but the gap goes through zero and the maxima are well observed. In principle, the $N-E$ effect should go to zero as the temperature goes to zero, but this is experimentally difficult to achieve because it requires a very small temperature gradient.

In order to understand and appreciate contributions and relative importance of various scattering modes in Pb$_{0.75}$Sn$_{0.25}$Se, which are used in the description of $N-E$ effect, we show in Fig. 3 an example of experimental mobility $\overline{\mu}$ for this sample, compared with the theory for the total mobility and partial mobilities related to single modes. The values of adjusted material parameters are given in Table 1. As indicated above, there is $\overline{\mu} = \mu(H)/A_r$, so that we have to go here through the calculation of $A_r$ as well. As the temperature goes from 0 to 300 K the calculated value of $A_r$ goes smoothly from 0.85 to 1.55. It can be seen that, with the indicated values of material parameters for various scattering modes, the overall description of mobility is quite satisfactory. It follows from the figure that the dominant scattering mechanism is due to alloy disorder. At higher temperatures the polar and nonpolar optical, as well as acoustical phonon modes become important. For the assumed low density of linear dislocations $N_{dis}$ = 10$^9\; {\rm cm}^{-2}$ the corresponding partial mobility is too high to be seen in the figure.

In Fig. 4 we plot theoretical behaviour of $P_{\rm N-E}$ for  Pb$_{1-x}$Sn$_x$Se having four different chemical   compositions $x$. All calculations are carried for small density of dislocations $N_{dis} = 1\cdot 10^8$ cm$^{-2}$, no damping ($G$ = 0) and fixed values of other material parameters. In addition, the electron density $N = 1\cdot 10^{18}$ cm$^{-3}$ is kept the same for all samples. This somewhat hypothetical calculation is instructive for several reasons. It shows that for $x$ = 0 and $x$ = 0.15, when no $E_g = 0$ occurs, the behaviour of $P_{\rm N-E}$ is very flat. On the other hand, for higher values of $x$, where the gap vanishes as a function of $T$, the $N-E$ effect has a maximum at the critical temperature $T_c$.
It is seen that, when no damping is introduced, theoretical $P_{\rm N-E}$ coefficients have high and very sharp peaks not observed experimentally. This, as we explained previously, is a result of the "explosive" behaviour of the $<{\mu}^2>$ integral at E$_g$ = 0, which appears in formula (10) for $P_{\rm N-E}$.

Figure 5 shows two calculations of the $N-E$ coefficient for the Pb$_{0.75}$Sn$_{0.25}$Se sample, carried out with the same values of adjustable material parameters. The dashed line indicates a calculation without damping, i.e. with $G$ = 0. It gives a sharp peak at the temperature $T_c$ = 141 K at which $E_g$ = 0. This peak is related to the integral $\overline{{\mu}^2}$ discussed above and it is not observed experimentally. For this reason we use the damping procedure mentioned in the preceding section with the adjusted value of $G$ = 20 meV. This gives the result indicated by the solid line. It is seen that now there is no sharp peak and the theory describes quite well the experiment shown in Fig. 1 with the exception of low temperatures.  Finally, we indicate  a marked difference of theoretical results for $P_{\rm N-E}$ calculated for the sample $x$ = 0.25 with the use of $G$ = 0, as seen in Figs 4 and 5.
This difference is due to largely different dislocation densities assumed for both calculations: $N_{dis} = 10^8\;{\rm cm}^{-2}$ in Fig. 4 and $N_{dis} = 10^9\;{\rm cm}^{-2}$ in Fig. 5.

Finally, Fig. 6 shows theoretical dependence of $P_{\rm N-E}$ versus temperature, calculated for all four Pb$_{1-x}$Sn$_x$Se samples of our interest. In the description we use for all samples the same values of material parameters, excepting the density of defects, the density of dislocations, and the damping constant $G$. The latter are adjusted for each sample and their values are used also in the description of average mobilities, as illustrated for the sample with $x$ = 0.25 in Fig. 3. In the same figure we reproduce for comparison the experimental data shown in Fig. 1 with the same vertical arrows indicating the critical temperatures $T_c$ at which $E_g$ = 0. The vertical arrows  drawn at maxima of P$_{\rm N-E}$ denote the critical temperature  $T_c$ at which E$_g$=0. The same arrows are transferred to Fig.1 for comparison with experimental data. It is seen that the calculated and experimental maxima occur at the same temperatures within several percent accuracy.
The general experimental and theoretical temperature behaviour near the maxima as well as the absolute values of P$_{\rm N-E}$ agree quite with each other. By comparing Figs. 4 and 5 it becomes clear that the maximum of $N-E$ effect is directly related to the gap going through zero. This is a general feature of the theory which should find applications for other zero-gap semiconductor and semimetal systems.
On the other hand, the low temperature behaviour of the samples with x=0.325 and x=0.39 remains unexplained according to the presented theory. It should be observed that the unexplained region corresponds to the nontrivial band ordering in which the topologically protected surface states appear, 
so we deal with a two-channel transport.
According to the formulas for two-channel mixed conductivity, the $N-E$ coefficient depends on weighted contributions of both conducting channels and difference of their thermoelectric powers \cite{Putley}. Consequently, it is possible that the low temperature behaviour is modified by the topological surface states which are not accounted for in our theory.   
It should be mentioned that higher values of the damping constant $G$ result in wider and lower theoretical maxima of $P_{\rm N-E}$. All in all, we achieve a good description of the $N-E$ effect and electron mobility for four investigated samples in the critical range of small forbidden gaps in which the transition between trivial-nontrivial band ordering takes place. Our analysis has general significance for zero-gap systems and, in particular, it should be useful for a description of electron transport in topological crystalline insulators and in recently discovered three-dimensional topological Dirac semimetals - bulk analogues of graphene \cite{lius,liun}.

\section{CONCLUSIONS AND SUMMARY}
The main conclusion of our work is that a maximum of the electron Nernst-Ettingshausen effect, measured as a function of decreasing temperature in Pb$_{1-x}$Sn$_x$Se alloys, indicates a critical temperature $T_c$ corresponding to the zero energy gap and transition from trivial to nontrivial band ordering in terms of the topological crystalline insulators. The above property allows one to determine experimentally critical temperatures $T_c$ by means of transport data. 
In particular, we checked theoretically that the maximum of $N-E$ effect does not depend on the carrier density in the sample.
Electron mobility and $N-E$ effect are measured and successfully described for four Pb$_{1-x}$Sn$_x$Se samples with 0.25$\leq x \leq$0.39 in the vicinity of $E_g = 0 $, establishing the dominant scattering modes. It is demonstrated that the description of Hall scattering factor and $N-E$ effect presents theoretical problems as the gap goes to zero and the bands become linear in the wave vector. The latter have been overcome introducing damping procedure. 
We emphasize that our conclusion concerning the maximum of $N-E$ effect is a result of the detailed theoretical treatment.
Our analysis of electron transport phenomena in the proximity of vanishing energy gap should be of value for other temperature and composition driven topological crystalline insulators and three-dimensional Dirac semimetals.

\section*{Acknowledgment}
We acknowledge support from Polish National Science Centre grants No. 2011/03/B/ST3/02659 and No. DEC-2012/07/B/ST3/03607, and from European Regional Development Fund through the Innovative Economy grant (POIG.01.01.02-00-108/09).

\newpage

\begin{table}
\caption
{Carrier densities and mobilities, as determined from Hall measurements of Pb$_{1-x}$Sn$_x$Se alloys and material parameters  used in the calculations. Scattering parameters are adjusted and kept the same for all samples. For alloy disorder: $U^c_{ad}$= -1.25$\cdot10^{-22}$ eV cm$^3$, $U^v_{ad}/U^c_{ad}$= -0.477, for short range potential A =1.6$\cdot10^{-21}$ eV cm$^3$, for acoustic deformation potential $E^c_{ac}$ = 35 eV, for optic nonpolar deformation potential: $E^c_{np}$ = 60 eV. Dislocation density $N_{dis}$ and damping constant $G$ are adjusted for each sample.}
\begin{ruledtabular}
\begin{tabular}{ccccc}
&x=0.25&x=0.277&x=0.325&x=0.39\\
&n-type&n-type & n-type& p-type \\
\hline
carrier density at 4.2K (cm$^{-3}$) & $3.1\cdot10^{18}$ &$3.6\cdot10^{18}$  &  $1\cdot10^{18}$ &  $2.7\cdot10^{18}$ \\
mobility at 4.2K   (cm$^{2}$/Vs) & 9300&5000 & 4500 & 9100 \\
carrier density at 300K (cm$^{-3}$) &$2.4\cdot10^{18}$ &$ 2.7\cdot10^{18}$&  $1.9\cdot10^{18}$& $2.2\cdot10^{18}$  \\
mobility at 300K   (cm$^{2}$/Vs) &1100 &1100 & 1200&1600 \\
thermopower at 300K ($\mu \rm V/K$)&-192&-180&-162&94 \\ \hline
 $G$(meV)& 20 & 10 & 5 & 8 \\
 $N_{dis}$(cm$^{-2}$) & $1\cdot10^9$ & $1.25\cdot10^{10}$ & $1.75\cdot10^{10}$ & $2\cdot10^{11}$ \\
\end{tabular}
\end{ruledtabular}
\end{table}
\ \\
\ \\
\begin{figure}
\includegraphics[scale=1.45,angle=0]{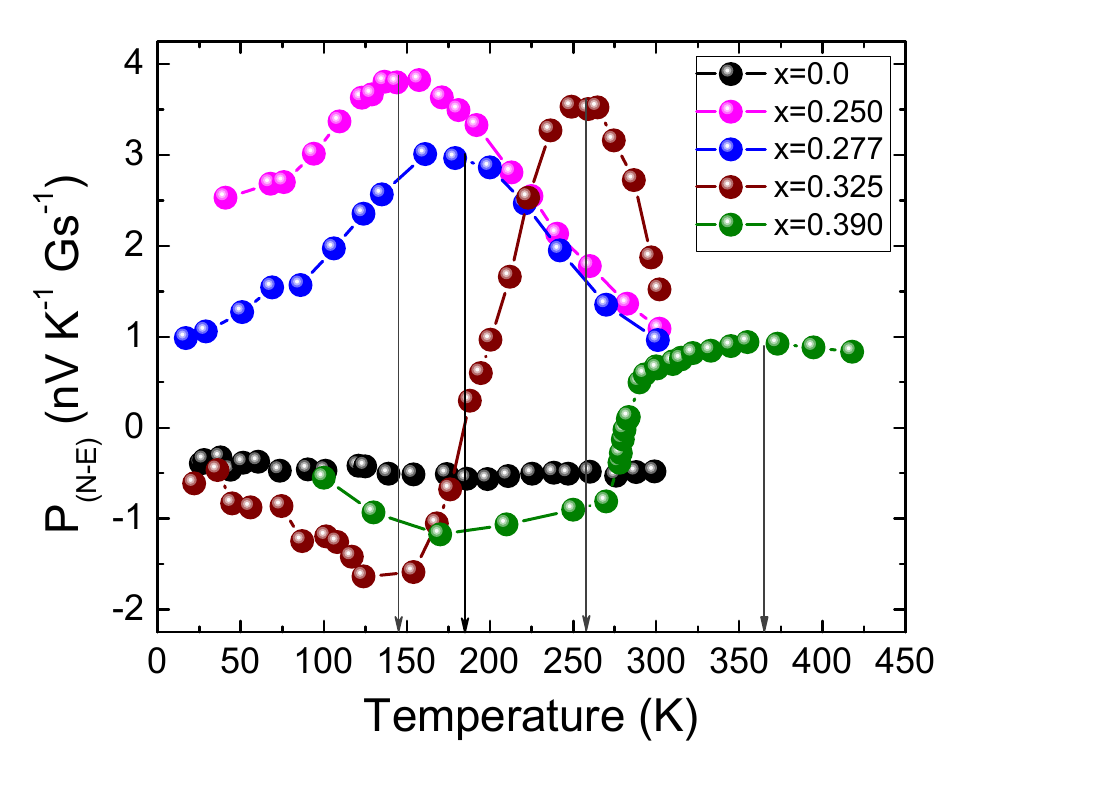}
\caption{Experimental values of the transverse $N-E$ coefficient, P$_{\rm N-E}$, versus temperature for PbSe and four samples of Pb$_{1-x}$Sn$_x$Se with different chemical compositions. Thin vertical arrows point to critical temperatures $T_c$ for which the band gap vanishes according to Eq. (4). It is seen that P$_{\rm N-E}$ has a maximum when the gap vanishes.}
\end{figure}

\begin{figure}
\includegraphics[scale=0.5,angle=0]{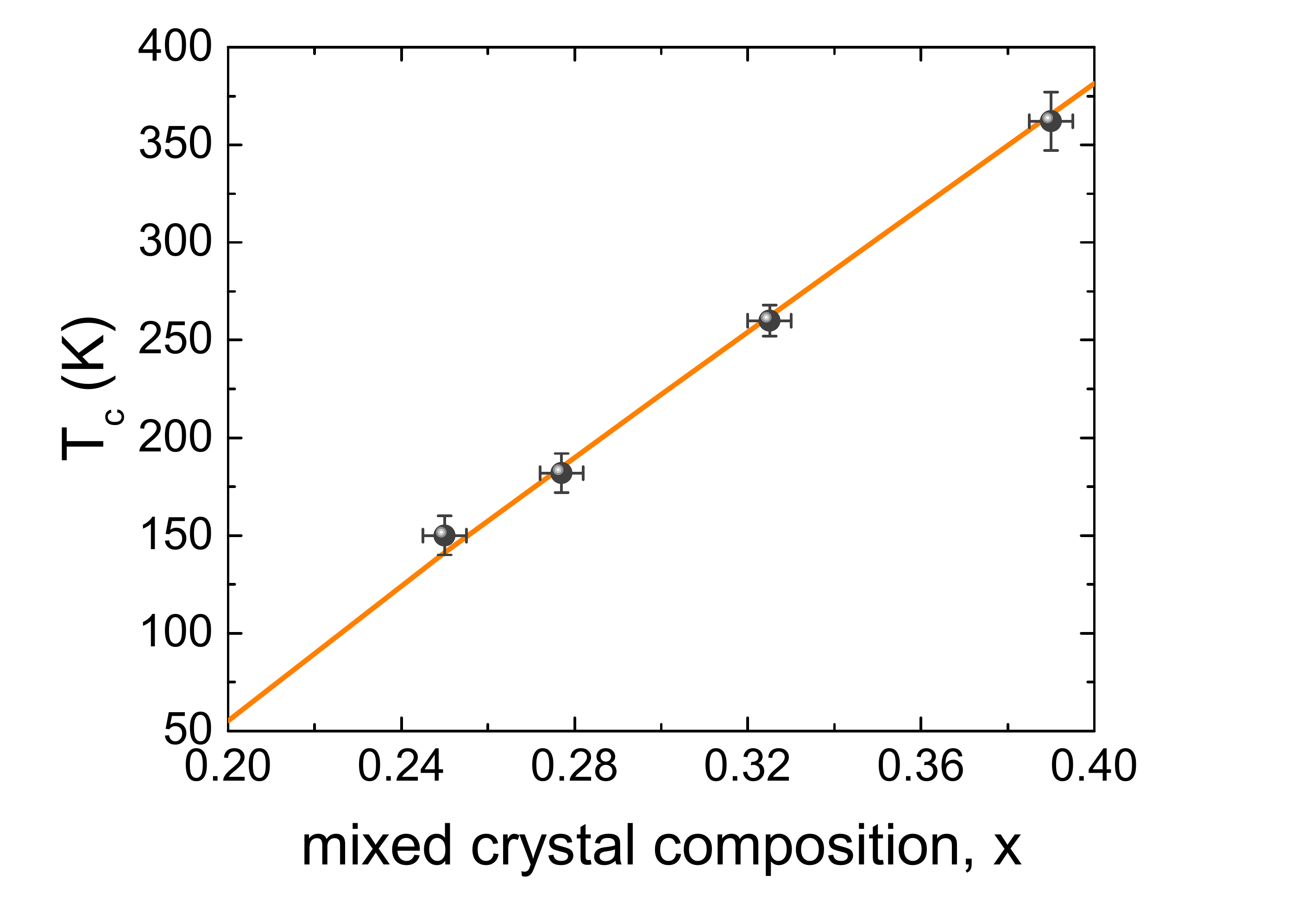}
\caption{Experimental temperatures of the maxima of $P_{\rm N-E}$ (T$_c$) taken from Fig. 1, for four investigated samples versus chemical composition $x$, compared with the line drawn according to Eq. (4).}
\end{figure}

\begin{figure}
\includegraphics[scale=0.5,angle=0]{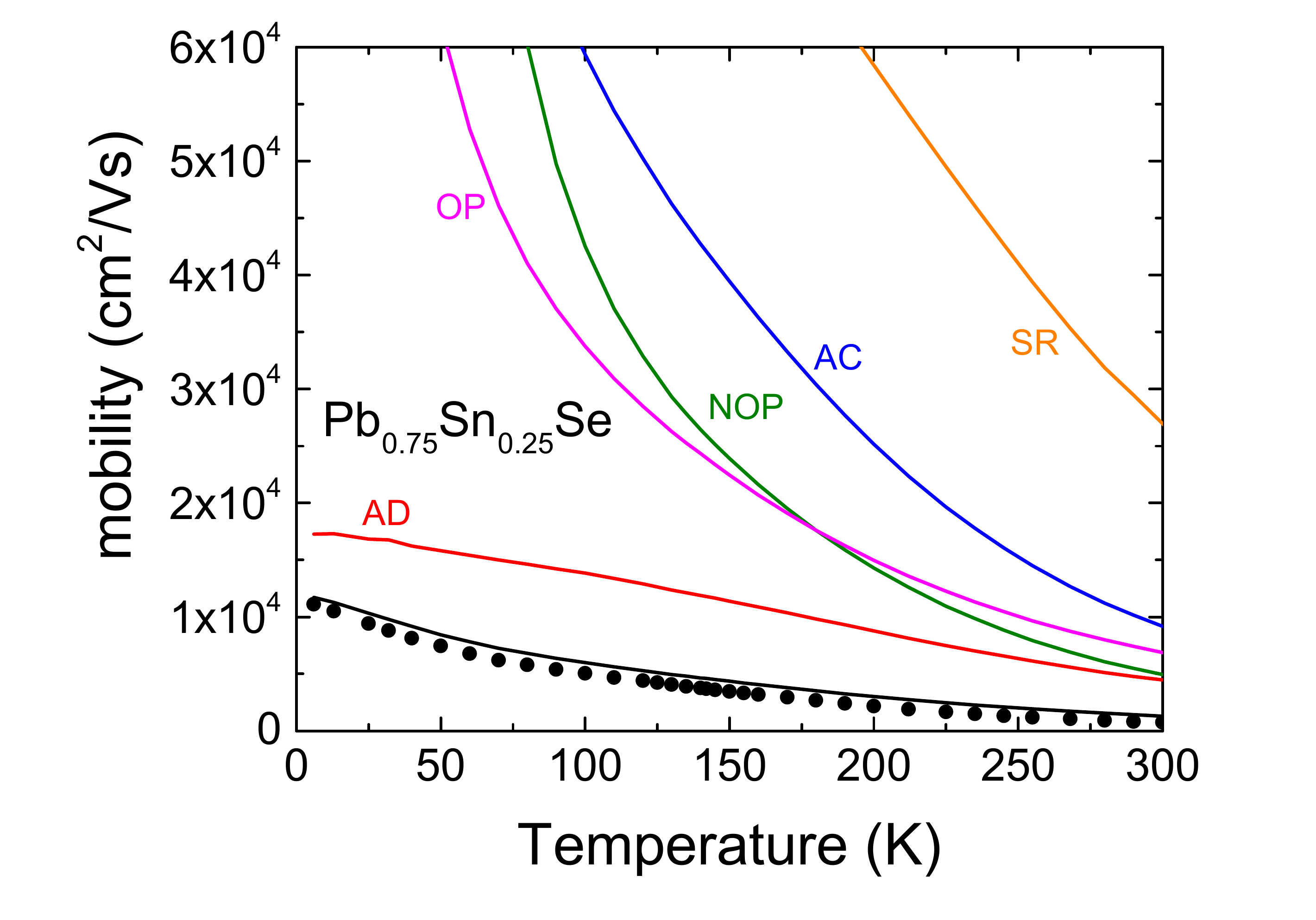}
\caption{Electron mobility $\overline{\mu}$ versus temperature for Pb$_{0.75}$Sn$_{0.25}$Se sample. Experiment - dots, theory - solid line. Theoretical partial mobilities related to separate scattering modes are also indicated. Notation: AD - alloy disorder, OP - polar optical phonons, NOP- nonpolar optical phonons, AC - acoustic phonons, SR - short range potential. Employed material parameters are given in Table 1.}
\end{figure}

\begin{figure}
\includegraphics[scale=0.5,angle=0]{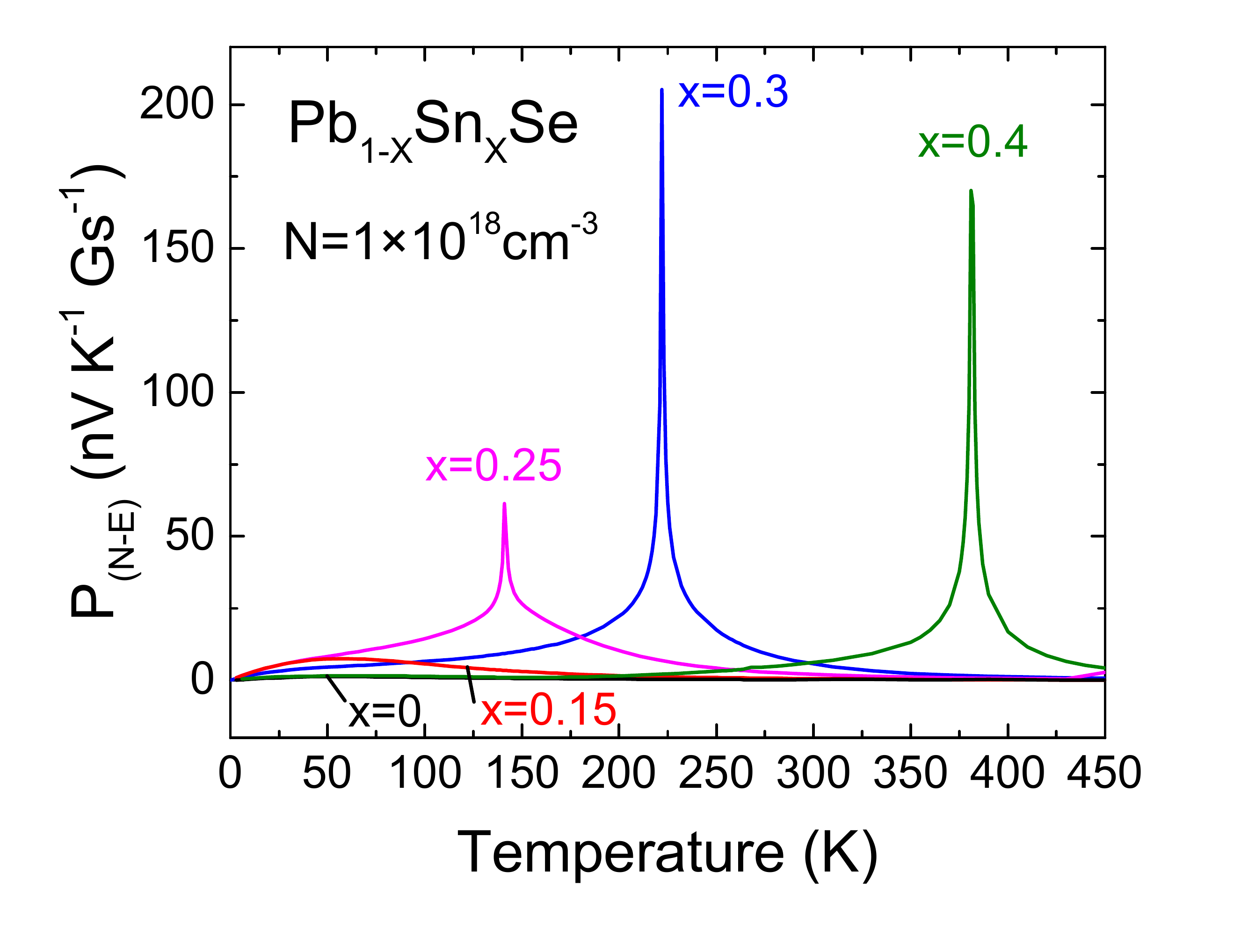}
\caption{Calculated $P_{\rm N-E}$ coefficient for PbSe and four Pb$_{1-x}$Sn$_x$Se samples versus temperature keeping the same material parameters, the same electron density $N$ and excluding damping. The assumed density of dislocations $N_{dis} = 10^{8}\;{\rm cm}^{-2}$ is very small. For zero damping, the $P_{\rm N-E}$ coefficients have sharp and narrow maxima at the critical temperatures $T_c$ for which the gaps vanish.}
\end{figure}

\begin{figure}
\includegraphics[scale=0.5,angle=0]{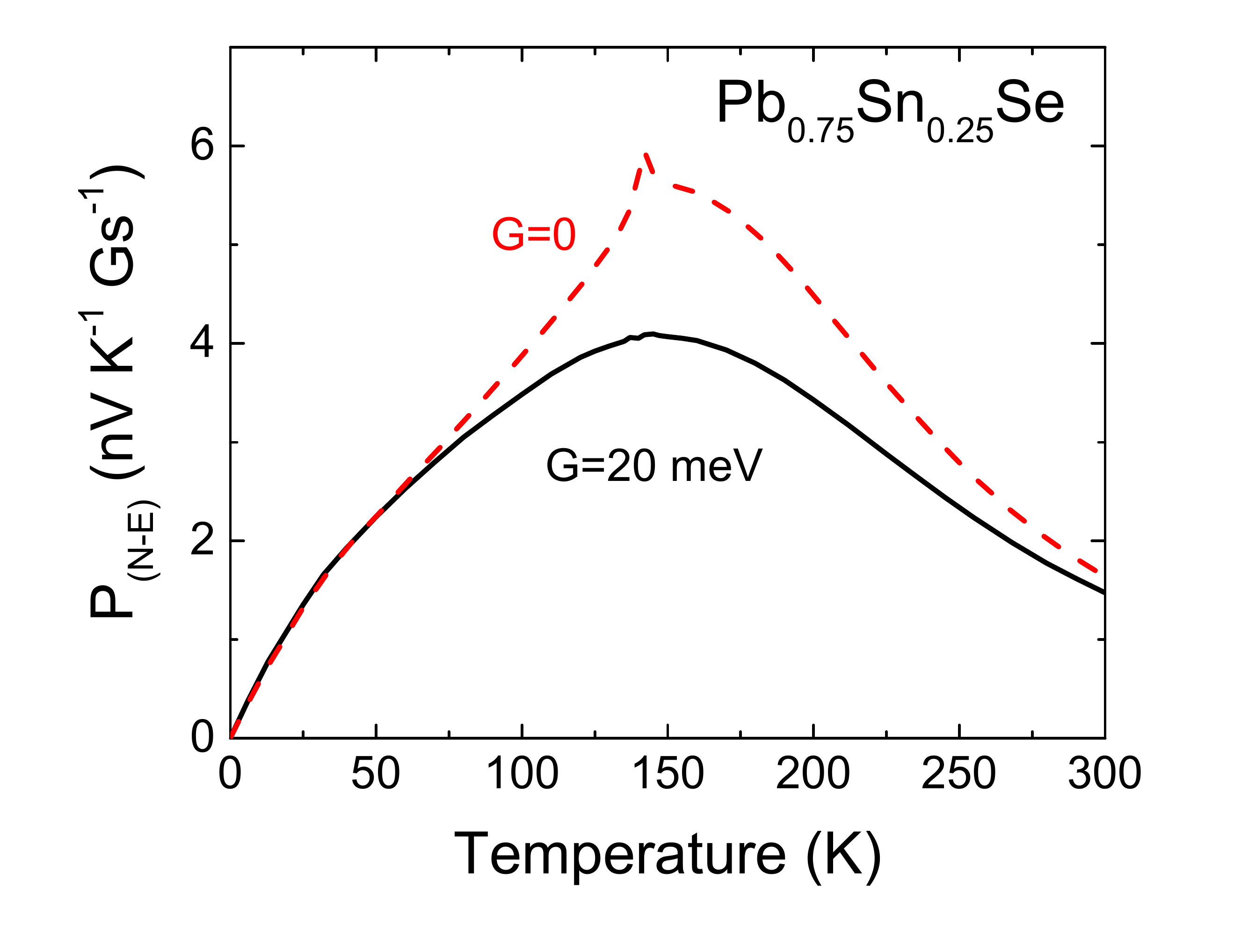}
\caption{The $N-E$ coefficient versus temperature calculated for Pb$_{0.75}$Sn$_{0.25}$Se sample. Material parameters are given in Table 1.  The dashed line is computed without damping, i.e. with $G$ = 0. The solid line includes damping ($G$ = 20 meV).}
\end{figure}

\begin{figure}
\includegraphics[scale=1.45,angle=0]{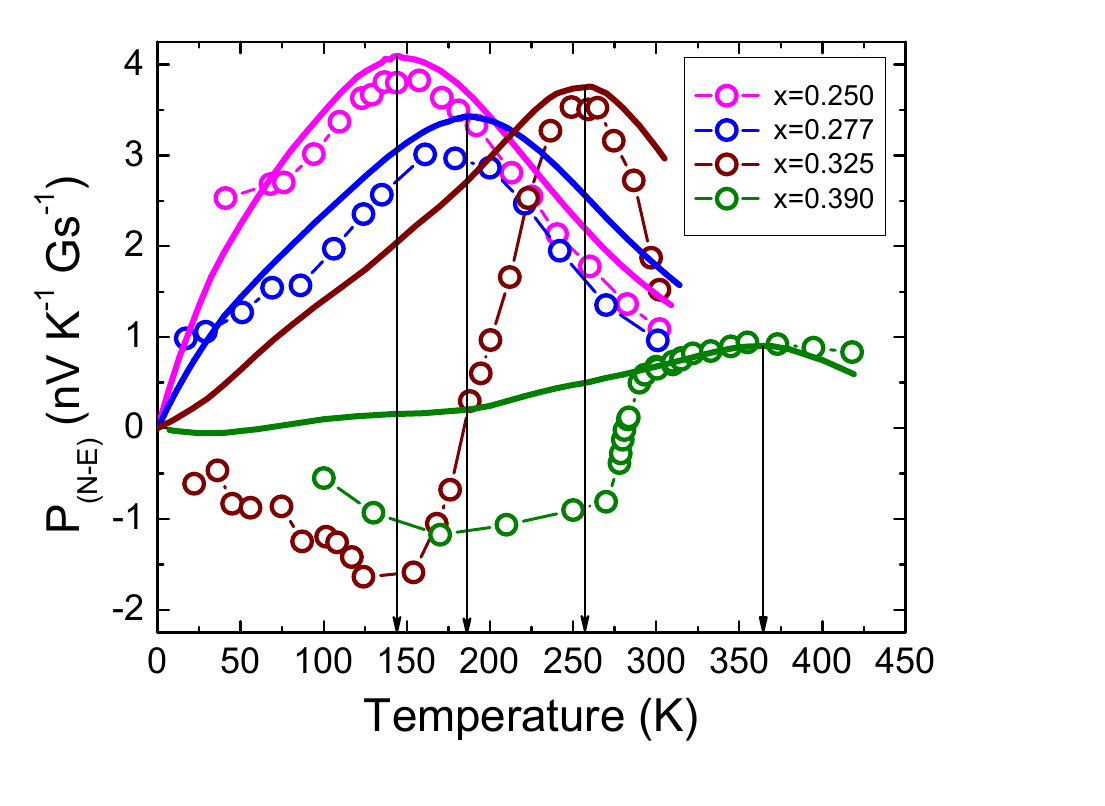}
\caption{Theoretical and experimental $P_{\rm N-E}$ coefficients versus temperature for four investigated Pb$_{1-x}$Sn$_x$Se samples.  Thick solid lines - theory with the use of material parameters given in Table 1, open circles - experiment. Thin vertical arrows are drawn at the critical temperatures $T_c$ for which the corresponding gaps vanish (the same arrows are also drawn in Fig. 1). For temperatures near maxima the theory describes very well the experimental data both in terms of shape and absolute values.}
\end{figure}

\end{document}